\documentclass[a4paper]{article}

\usepackage[nolist,nohyperlinks]{acronym}
\usepackage{INTERSPEECH2022}
\ninept

\usepackage{amssymb}
\usepackage{pifont}%
\usepackage{algorithm,setspace}
\usepackage[noend]{algpseudocode}
\usepackage{comment}

\usepackage{adjustbox}
\usepackage{tikz}
\usepackage{pgfplots}
    \usetikzlibrary{calc,positioning,patterns}
    \pgfplotsset{compat=1.3}
\usepackage{pgfplotstable}
\usepackage{booktabs}

\newcommand{\cgauss}[2]{\mathcal{N}_{\mathbb{C}}({#1}; {#2})}
\newcommand{\C}[0]{\mathbb{C}}

\newcommand{\x}[1]{\boldsymbol{{#1}}_{t, f}}
\newcommand{\est}[1]{\widehat{\boldsymbol{{#1}}}_{t, f}}
\newcommand{\xt}[2]{\boldsymbol{{#1}}_{#2}}

\newcommand{\phie}[0]{\phi^{(q)}}

\newcommand{\T}[0]{\mathrm{T}_\mathrm{60}}

\newcommand\blfootnote[1]{%
  \begingroup
  \renewcommand\thefootnote{}\footnote{#1}%
  \addtocounter{footnote}{-1}%
  \endgroup
}

\title{Neural Network-augmented Kalman Filtering for Robust Online Speech Dereverberation in Noisy Reverberant Environments}

\name{Jean-Marie Lemercier$^{\star}$, Joachim Thiemann$^{\dagger}$, Raphael Koning$^{\dagger}$, Timo Gerkmann$^{\star}$}
\address{$^{\star}$Signal Processing (SP), Universit\"at Hamburg, Germany\\ $^{\dagger}$Advanced Bionics, Hanover, Germany}
\email{\{firstname.lastname\}@uni-hamburg.de, \{firstname.lastname\}@advancedbionics.com}

\begin{document}

\maketitle
\begin{abstract}

In this paper, a neural network-augmented algorithm for noise-robust online dereverberation with a Kalman filtering variant of the weighted prediction error (WPE) method is proposed.
The filter stochastic variations are predicted by a deep neural network (DNN) trained end-to-end using the filter residual error and signal characteristics.
The presented framework allows for robust dereverberation on a single-channel noisy reverberant dataset similar to WHAMR!. 

The Kalman filtering WPE introduces distortions in the enhanced signal when predicting the filter variations from the residual error only, if the target speech power spectral density is not perfectly known and the observation is noisy.
The proposed approach avoids these distortions by correcting the filter variations estimation in a data-driven way, increasing the robustness of the method to noisy scenarios. 
Furthermore, it yields a strong dereverberation and denoising performance  compared to a DNN-supported recursive least squares variant of WPE, especially for highly noisy inputs.

\end{abstract}

\blfootnote{This work has been funded by the Federal Ministry for Economic Affairs and Climate Action, project 01MK20012S, AP380. The authors are responsible for the content of this paper.}

\section{Introduction}
\label{sec:intro}

Communication and hearing devices require modules aiming at suppressing undesired parts of the signal to improve the speech characteristics. Amongst these is reverberation caused by room acoustics, where late reflections particularly degrade the speech quality and intelligibility \cite{Naylor2011}. Presence of additional background noise and interfering speakers further worsens the ability to clearly perceive target speech.

In complement to traditional single-channel schemes, many multi-channel algorithms leveraging spatial and spectral information were proposed for enhancement of noisy reverberant speech.
Traditional approaches include beamforming \cite{Doclo2005, Kuklasinski2015, Leglaive2016}, possibly combined with spectral enhancement \cite{Cauchi2015}, coherence-weighting \cite{Gerkmann2011, Schwarz2015}, and \ac{MCLP} based approaches such as the well-known \ac{WPE} algorithm \cite{Nakatani2008b, Jukic2015}.
\ac{WPE} computes an auto-regressive multi-channel filter in the short-time spectrum and applies it to a delayed group of reverberant speech frames. It requires an estimate of the target speech \ac{PSD}, estimated either by statistical models \cite{Nakatani2008b, Jukic2015} or \acp{DNN} \cite{Kinoshita2017, Heymann2018}.

In order to cope with real-time requirements and changing acoustics, online adaptive dereverberation methods derived from \ac{MCLP} approaches were introduced. These methods are based on either \ac{KF} \cite{Schwartz2015, Braun2016, Braun2018, Hashem2020, Braun2021} or on a \ac{RLS} adapted \ac{WPE}, which can be seen a special case of \ac{KF} \cite{Yoshioka2009, Caroselli2017, Heymann2018, Lemercier2022}.
Online convolutional beamformers performing joint dereverberation and denoising based on either \ac{RLS} or \ac{KF} were proposed in \cite{Nakatani2019, Dietzen2020, Dietzen2017, Hashem2020, Braun2021}.

Kalman filtering \ac{WPE} (KF-WPE) is a particularly interesting framework for adaptive dereveberation in noisy and dynamic environments. It updates the filter in a much faster and more flexible way than its \ac{RLS} counterpart, due to a Gauss-Markov model of the filter transition dynamics \cite{Haykin2001}.
However, it is known that the Kalman filter can be particularly sensitive to estimation errors in the unobservable state space model parameters \cite{Anderson1979}. In KF-WPE, these parameters include the target speech \ac{PSD} and the filter transition covariance.

The contributions of this work are threefold.
First, we introduce a low-complexity variant of KF-WPE based on \cite{Braun2016} and a scaled identity model of the speech covariance matrix.
Secondly, we evaluate the performance and robustness of this KF-WPE variant in noisy reverberant environments. 
Finally, we present an estimation strategy of both the target speech \ac{PSD} and the filter transition covariance based on \acp{DNN}.
In this strategy, a first DNN estimates the target speech \ac{PSD} from the noisy reverberant mixture. 
A second DNN then estimates the filter transition covariance from the filter residual error and signal characteristics, and is trained with an end-to-end criterion.

This framework enables a robust estimation of the filter transition covariance under target \ac{PSD} uncertainty. The resulting DNN-augmented KF-WPE performs stronger dereverberation and denoising than RLS-WPE, with less distortions than traditional KF-WPE --- all approaches using the same DNN-supported target PSD estimator. The approach is robust to noisy observations and even removes a lot of environmental noise, although noise is not accounted for in the signal model.

The rest of this paper is organized as follows.  In Section \ref{sec:kwpe}, the low-complexity variant of KF-WPE is summarized.
Section \ref{sec:varnet} presents the DNN-supported strategy to predict the \ac{WPE} filter transition covariance and the target speech \ac{PSD}.
In Section \ref{sec:exp}, we describe the experimental setup and training strategy. Results are presented and discussed in Section \ref{sec:results}.

\section{Kalman filtering adapted \ac{WPE} dereverberation}
\label{sec:kwpe}

\subsection{Signal model}

In the \ac{STFT} domain using the subband-filtering approximation \cite{Nakatani2008b}, the noisy reverberant speech $\boldsymbol{x} = [x^{(1)}, \dots, x^{(D)}] \in \C^{D}$ is obtained at the microphone array by convolution of the anechoic speech $s$ and the \acp{RIR} $\boldsymbol{H} \in \C^{D \times D}$ with length $N$:
\vspace{-0.2cm}
\begin{align} \label{eq:signal_model0}
    \x{x} &= \displaystyle{\sum_{\tau = 0 }^{N}}\boldsymbol{H}_{\tau, f} s_{t-\tau, f} + \x{n} \\
    &= \x{d} + \x{e} + \x{r} + \x{n},
\end{align}

\noindent where $t$ denotes the time frame index and $f$ the frequency bin, which we will drop when not needed. 
$\boldsymbol{d}$ denotes the direct path, $\boldsymbol{e}$ the early reflections component,  $\boldsymbol{r}$ the late reverberation and $\boldsymbol{n}$ an error term comprising modelling errors and environmental noise. The early reflections $\boldsymbol{e}$ were shown to contribute to speech quality and intelligibility for normal and hearing-aided listeners \cite{Bradley2003}. The dereverberation objective is therefore to retrieve $\boldsymbol{\nu} = \boldsymbol{d + e}$.

However, early reflections may be detrimental for some people, e.g. cochlear-implant users,  particularly in highly-reverberant scenarios \cite{Hu2014}. Accordingly, the dereverberation objective can be adjusted to different listener categories \cite{Lemercier2022}. %

\subsection{WPE Dereverberation}

As in \cite{Nakatani2008b}, the anechoic speech $s$ is modelled in the \ac{STFT} domain with a zero-mean time-varying Gaussian model. It follows from \eqref{eq:signal_model0} that the target speech also is a zero-mean time-varying Gaussian process with time-frequency dependent covariance:
\begin{equation} \x{\nu} \sim \cgauss{0}{\boldsymbol{\Phi}^{(\nu)}_{t,f}} \end{equation}

The \ac{WPE} algorithm \cite{Nakatani2008b} uses an auto-regressive model to approximate the late reverberation $\boldsymbol{r}$. A multi-channel filter $\boldsymbol{G} \in \C^{D^2K}$ with $K$ taps is estimated, aiming at representing the inverse of the late tail of the RIRs $\boldsymbol{H}$. 
The target $\boldsymbol{\nu}$ is then obtained through linear prediction, with a delay $\Delta$ avoiding undesired short-time speech cancellations, which also leads to preserving parts of the early reflections. We omit the frequency index $f$ in the following as computations are performed in each frequency band independently. 
By disregarding the error term $\boldsymbol{n}$ in \eqref{eq:signal_model0} in noiseless scenarios we obtain:
\begin{equation} \label{eq:MCDLP}
    \boldsymbol{\nu}^{(\mathrm{WPE})}_{t} = \boldsymbol{x}_{t} -  \boldsymbol{G}^H ( \boldsymbol{I}_D \otimes \boldsymbol{X}_{t - \Delta} ) ,
\end{equation}
where $\boldsymbol{X}_{t - \Delta} = \begin{bmatrix} \boldsymbol{x}^T_{t-\Delta}, \dots, \boldsymbol{x}^T_{t-\Delta-K+1} \end{bmatrix}^T \in \C^{DK}$ and $\otimes$ is the Kroenecker product.

\subsection{Kalman Filtering WPE Dereverberation}

In order to obtain an adaptive and real-time capable approach, a \ac{KF} variant of \ac{WPE} was proposed in \cite{Braun2016}, where the \ac{WPE} filter $\boldsymbol{G}$ is recursively updated with a Markov model:
\begin{equation} \boldsymbol{G}_{t} = \boldsymbol{G}_{t-1}  + \boldsymbol{q}_t , \label{eq:markov} \end{equation}
where $\boldsymbol{q}$ is the filter transition stochastic noise following a zero-mean Gaussian process $\boldsymbol{q}_t \sim \cgauss{0}{\boldsymbol{Q}_t}$.

A Gaussian-Markov state-space model is formed by the transition equation \eqref{eq:markov} and the observation model \eqref{eq:MCDLP}, where $\boldsymbol{G}$ is the state, $\boldsymbol{x}$ the observation and $\boldsymbol{\nu}^{(\mathrm{WPE})}$ the \textit{observation noise}. Given these and some independency assumptions described in \cite{Braun2016}, the Kalman filter $\widehat{\boldsymbol{G}}$ is an optimal recursive estimator with respect to the mean-squared error criterion:
\begin{equation} \arg \underset{\boldsymbol{G}}{\min} \, \mathbb{E}_t \{ \, || \boldsymbol{G}_t - \boldsymbol{\widehat{G}}_t  ||_2^2 \, \} , \end{equation}

Similarly to \cite{Nakatani2008b, Yoshioka2009}, we make the further assumption that the target speech $\boldsymbol{\nu}$ is modelled identically and independently at each microphone, thus making the target speech covariance a scaled identity matrix characterized by \ac{PSD} $\lambda$:
\begin{equation} \boldsymbol{\Phi}^{(\nu)}_{t} = \lambda_{t} \boldsymbol{I}_{D} \label{eq:model_cov} \end{equation}

The \ac{WPE} filter $\boldsymbol{G}^{(d)} \in \C^{DK}$ is then estimated and applied for each channel $d$ independently, which considerably curbs the algorithmic complexity in a multi-channel setting.

We define the filter error covariance matrix as:
\begin{equation} \boldsymbol{\Phi}^{(\epsilon)} = \mathbb{E}_{t,d} \{ \, [ \boldsymbol{G}_t^{(d)} - \boldsymbol{\widehat{G}}_t^{(d)}  ] [ \boldsymbol{G}_t^{(d)} - \boldsymbol{\widehat{G}}_t^{(d)}  ]^H \}
\end{equation}

The corresponding derivations are adapted from \cite{Braun2016}:
\begin{equation} \label{eq:apriori}
    \boldsymbol{\Phi}^{(\epsilon)}_{t|t-1} =  \boldsymbol{\Phi}^{(\epsilon)}_{t-1|t-1} + \boldsymbol{Q}_{t-1} ,
\end{equation}
\begin{equation} \label{eq:kalman} \xt{K}{t} = \displaystyle{\frac{
    \boldsymbol{\Phi}^{(\epsilon)}_{t|t-1} \boldsymbol{X}_{t - \Delta} }{
    \lambda_{t} + \boldsymbol{X}^H_{t - \Delta} \boldsymbol{\Phi}^{(\epsilon)}_{t|t-1} \boldsymbol{X}_{t - \Delta} }} , \end{equation}
\begin{equation} \label{eq:invcov} \boldsymbol{\Phi}^{(\epsilon)}_{t|t} = \boldsymbol{\Phi}^{(\epsilon)}_{t|t-1} - \xt{K}{t} \boldsymbol{X}^T_{t-\Delta} \boldsymbol{\Phi}^{(\epsilon)}_{t|t-1} , \end{equation}
\begin{equation} \label{eq:filter} \boldsymbol{\widehat{G}}^{(d)}_{t} = \boldsymbol{\widehat{G}}^{(d)}_{t-1}  + \xt{K}{t} \big( x^{(d)}_{t} - ( \boldsymbol{\widehat{G}}^{(d)}_{t-1} )^H  \boldsymbol{X}_{t-\Delta, f} \big)^H , \end{equation}
\begin{equation}
\label{eq:MCDLP_simplified}
\widehat{\nu}_t^{(d)} = x_t^{(d)} -  ( \boldsymbol{\widehat{G}}^{(d)}_{t} )^H  \boldsymbol{X}_{t-\Delta, f} .    
\end{equation}

\noindent where $\boldsymbol{K} \in \C^{DK}$ is the Kalman gain.

\subsection{\ac{RLS}-WPE dereverberation}

A \ac{RLS} version of \ac{WPE} was introduced in \cite{Yoshioka2009} and is equivalent to KF-WPE for the static case, i.e. when $\boldsymbol{q}=0$.
The target speech covariance is replaced by a scaled identity matrix as in \eqref{eq:model_cov}. The filter error covariance is thus equivalent to the inverse of the weighted covariance of the reverberant buffer:
\begin{equation} \boldsymbol{\Phi}^{(\epsilon)}_\mathrm{RLS} = \left[ \mathbb{E}_t \left\{ \displaystyle{\frac{\boldsymbol{X}_{t - \Delta} \boldsymbol{X}_{t - \Delta}^H }{ \lambda_t }} \right\} \right]^{-1} . \end{equation}

The computations are equivalent to \eqref{eq:apriori}-\eqref{eq:filter}, only that the filter error covariance is updated recursively with forgetting factor $\alpha$ \cite{Haykin2001}, replacing \eqref{eq:apriori} with:
\begin{equation} \label{eq:apriori_RLS}
        \boldsymbol{\Phi}^{(\epsilon)}_{\mathrm{RLS}, t|t-1} =  \frac{1}{\alpha} \boldsymbol{\Phi}^{(\epsilon)}_{\mathrm{RLS}, t-1|t-1} .
\end{equation}
Processing order is then \eqref{eq:apriori_RLS}$\xrightarrow{}$\eqref{eq:kalman}$\xrightarrow{}$\eqref{eq:invcov}$\xrightarrow{}$\eqref{eq:filter}$\xrightarrow{}$\eqref{eq:MCDLP_simplified}.

\section{Filter transition covariance estimation}
\label{sec:varnet}

\subsection{Covariance model}

The filter transition covariance $\boldsymbol{Q}$ is an unknown parameter which must be estimated at each step to update the filter error covariance $\boldsymbol{\Phi}^{(\epsilon)}$ in \eqref{eq:apriori}. As in \cite{Braun2016} we model each filter tap transition identically and independently, which results in $\boldsymbol{Q}$ being an identity matrix scaled by the \textit{filter transition power} $\phie$:
\begin{equation} \label{eq:cov_model}
    \boldsymbol{Q}_t = \phie_t \boldsymbol{I}_{DK} .
\end{equation}

We furthermore define the \textit{filter residual error} $e$ as:
\begin{equation}
   e_t = \mathbb{E}_d \{ \, || \boldsymbol{G}^{(d)}_{t} - \boldsymbol{G}^{(d)}_{t-1} ||_2^2 \, \} .
   \label{eq:error}
\end{equation}

In \cite{Braun2016}, the filter transition power is simply modelled by adding a small fixed parameter $\eta$ to the scaled residual error $e$, in order to force a permanent adaptation even if the filter did not vary at the previous time step:
\begin{equation}
\label{eq:phi_braun}
    \phie_t = \displaystyle{\frac{1}{D K}}e_t + \eta
\end{equation}

\subsection{DNN estimation}

As we will show in Section~\ref{sec:results}, using the filter residual error $e$ to model the filter transition power $\phie$ is a straightforward model which yields excellent results if the oracle target \ac{PSD} is available. However, if the \ac{PSD} estimation is flawed due to noisy observations and limited prediction power, using the same filter transition power model \eqref{eq:error} introduces problematic speech distortions. In particular, the bias $\eta$ should be adapted as a function of the input \ac{SNR}, decreasing the filter transition power if the observation is too uncertain.

Rather than using a statistical approach based on input \ac{SNR} analysis, we propose here a \ac{DNN} strategy to infer at each frame the filter transition power $\phie$ directly from data, using an end-to-end criterion to optimize the \ac{DNN} parameters. We argue that this increases the robustness of KF-WPE to erroneous \ac{PSD} estimation.
A related concept is reported in \cite{Haubner2022}, where a DNN is used to learn the step size, i.e. the Kalman gain, of a block-wise adaptive system identification algorithm. Our approach is however distinct as (i)- we perform frame-wise adaptive dereverberation in noisy environments, (ii)- we do not use a single \ac{DNN} inferring both step size parameters, but use separated networks with distinct inputs and training strategies; and (iii)- the criterion optimizes the \acp{DNN} with respect to the estimated signal, and not the filter error, which is not available in this task.

The \ac{DNN}---called here $\mathrm{VarNet}$---takes as its input at every step $t$ a vector containing the channel-averaged reverberant speech periodogram $|\bar{x}_t|^2 = \frac{1}{D} \sum_{d=1}^D |x_{t}^{(d)}|^2 \, \} $, the target speech PSD estimate $\widehat{\lambda}_{t}$ and the filter residual error $e_t$.
The filter transition power $\phie_t$ is then obtained on a model inspired by \eqref{eq:phi_braun}, where the $\mathrm{VarNet}$ positive real-valued output mask $\mathcal{M_t^{(\eta)}}$ is multiplied by a maximal bias value $\eta_\mathrm{max}$ before being added to the scaled filter residual error $e$:
\begin{equation}
    \phie_t = \displaystyle{\frac{1}{D K}} e_t + \eta_\mathrm{max} \odot \mathcal{M}^{(\eta)}_{t}(|\bar{x}_t|^2, \widehat{\lambda}_t, e_t)
    \label{eq:phi_dnn}
\end{equation}

The target speech \ac{PSD} estimate $\widehat{\lambda}_t$ is obtained from the channel-averaged magnitude $|\bar{x}_t|= \frac{1}{D} \sum_{d=1}^D |x_{t}^{(d)}| \, \} $ by a preceding \ac{DNN}---denoted as $\mathrm{MaskNet}$---as in \cite{Heymann2018, Lemercier2022}:
\begin{equation}
    \widehat{\lambda}_t = \big( \mathcal{M}^{(\nu)}_t \odot |\bar{x}_t| \big)^2,
\end{equation}
with $\mathcal{M}^{(\nu)}_t$ being the estimated real-valued positive mask.

\subsection{Training strategy}

The $\mathrm{MaskNet}$ is pre-trained with a mask-based objective:
\begin{equation}
\mathcal{L}^\mathrm{(pre)} = \mathcal{L} (\mathcal{M}^{(\nu)} \odot |\boldsymbol{x}| , |\boldsymbol{\nu}| ) ,
\label{eq:pre-training}
\end{equation}
with $\mathcal{L}$ being the chosen loss function. The parameters of $\mathrm{MaskNet}$ are then frozen, and the $\mathrm{VarNet}$ is trained with the end-to-end criterion:
\begin{equation}
\mathcal{L}^\mathrm{(end)} = \mathcal{L} ( | \widehat{\boldsymbol{\nu}} | , |\boldsymbol{\nu}| ) .
\label{eq:fine-tuning}
\end{equation}

\noindent Schematics of the algorithm are displayed in \figurename~\ref{fig:algo}.
\tikzstyle{sum} = [draw, fill=white, circle, node distance=1cm]
\tikzstyle{input} = [coordinate]
\tikzstyle{output} = [coordinate]
\tikzstyle{pinstyle} = [pin edge={to-,thin,black}]
\tikzstyle{branch}=[fill,shape=circle,minimum size=3pt,inner sep=0pt]
\tikzstyle{connarrow}=[-latex, line width=0.125em]
\tikzstyle{connline}=[-, line width=0.125em]
\tikzstyle{block} = [draw, rectangle, minimum height=3em, minimum width=3em, line width=1pt]
\tikzstyle{circ} = [draw, circle, radius=1em, minimum width=1em, line width=1pt]

\begin{figure}
    \hspace{-0.4cm}
    \tikzstyle{block} = [draw, rectangle, minimum height=3em, minimum width=3em, line width=1pt]
    \tikzstyle{circ} = [draw, circle, radius=1em, minimum width=1em, line width=1pt]
    \scalebox{0.71}{
    \begin{tikzpicture}[auto]
    
        \node at (2.3, 0.2) (input) {};
        \node[branch] at (3.0, 0.2) (branch_in_buffer) {};
        
        \node[block, align=center] at (4.0, 0.2) (D-delay) {$z^{-\Delta}$};
        \node[block] at (6.5, 0.2) (buffer) {Buffering};
        \node[branch] at (9.0, 0.2) (bufferbranchout1) {};

        \node[block] at (4.0, -3.9) (mag) {$| \cdot |$};
        \node[branch] at (5.1, -3.9) (branch_in_dnn) {};

        \node[block, fill=blue!14, align=center] at (6.5, -3.9) (masknet) {$\mathrm{MaskNet}$};
        
        \node[block] at (6.5, -5.2) (square2) {$\cdot^2$};

        \node[branch] at (7.6, -3.9) (branch_psd) {};
        
        \node[block, fill=yellow!14, align=center] at (9.0, -1.4) (kalman) {Kalman Gain\\Computation};
        \node[block, fill=yellow!14, align=center] at (11.8, -1.4) (wpe) {WPE Filter\\Computation};
        \node[block, fill=blue!14, align=center] at (9.0, -3.9) (varnet) {$\mathrm{VarNet}$};

        \node[block, fill=yellow!14] at (11.8, 0.2) (filter) {WPE Filtering};
        
        \node[block, align=center] at (11.8, -3.0) (1-delay) {$z^{-1}$};

        \node at (13.8, 0.2) (output) {};

        \draw[connline] (input) -- (branch_in_buffer);
        \draw[connarrow] (branch_in_buffer) --  node[pos=0.0, xshift=-0.07cm] {$\boldsymbol{x}_{t,f}$} (D-delay);
        \draw[connarrow] (D-delay) -- node[pos=0.5] {$\boldsymbol{x}_{t-\Delta,f}$} (buffer);
        \draw[connarrow] (branch_in_buffer) |- (mag);
        
        \draw[connline] (mag) --  node[pos=1.0,xshift=0.0cm] {$|\bar{x}_{t,f}|$} (branch_in_dnn);
        \draw[connarrow] (branch_in_dnn) -- (masknet);
        \draw[connline] (branch_in_dnn) |- (square2);
        \draw[connarrow] (square2) -| node[pos=0.27] {$|\bar{x}_{t,f}|^2$} (varnet.south);
        
        \draw[connline] (masknet) -- (branch_psd);
        \draw[connarrow] (branch_psd) |- (kalman);
        \draw[connarrow] (branch_psd) -- node[pos=0.5] {$\lambda_{t,f}$}  (varnet.west);
        
        \draw[connline] (buffer)  -- node[pos=1.0]  {$\boldsymbol{X}_{t-\Delta,f}$} (bufferbranchout1);
        \draw[connarrow] (bufferbranchout1) -- (kalman);
        \draw[connarrow] (bufferbranchout1) -- (filter);
        
        \draw[connarrow] (varnet) -- node[pos=0.45, right] {$\phie_{t,f}$} (kalman); 

        \draw[connarrow] (kalman) -- node[pos=0.5, above] {$\boldsymbol{K}_{t,f}$} (wpe);
        
        \draw[connarrow] (wpe.north) -- node[pos=0.5, right] {$\boldsymbol{G}_{t,f}$} (filter);
        \draw[connarrow] (wpe.south) -- node[pos=0.5, right] {$e_{t,f}$} (1-delay);
        
        \draw[connarrow] (1-delay) |- node[pos=0.5, right] {$e_{t-1,f}$} (varnet.east);

        \draw[connarrow] (filter) -- node[pos=0.5] {$\est{\nu}$}  (output);

    \end{tikzpicture}
    }
    \caption{\centering DNN-supported Kalman-filtering adapted dereverberation. Blue blocks refer to trainable \ac{DNN} layers. Yellow blocks represents adaptive statistical signal processing. $z^{-\tau}$ blocks implement a time delay of $\tau$ STFT frames.}
    \label{fig:algo}
    \vspace{-5mm}
\end{figure}
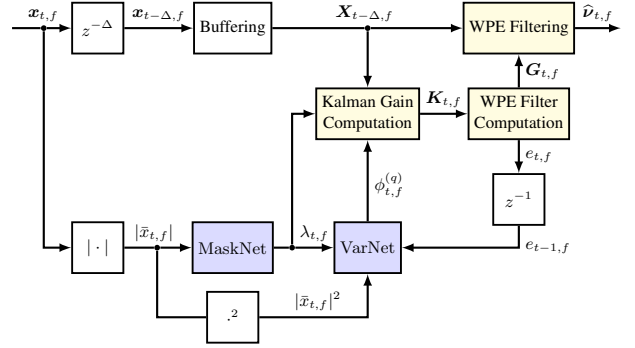

\section{Experimental setup}
\label{sec:exp}

\subsection{Dataset generation}

The data generation method resembles that of the WHAMR! dataset \cite{Maciejewski2020}. We concatenate anechoic speech utterances from the WSJ0 dataset belonging to the same speaker, and construct sequences of approximately $20$ seconds.
These sequences are convolved with $2$-channel RIRs generated with the RAZR engine \cite{Wendt2014a} and randomly picked. Each RIR is generated by uniformly sampling room acoustics parameters as in \cite{Maciejewski2020} and a $\text{T}_\text{60}$ reverberation time between $0.4$ and $1.0$ seconds. \ac{HRTF} auralization is performed in the RAZR engine, using a KEMAR dummy head response from the MMHR-HRTF database \cite{Thiemann2019}.
Finally, $2$-channel noise from the WHAM! recorded noise dataset \cite{Wichern2019} is added to the reverberant mixture with a \ac{SNR}---relative to the reverberant signal---uniformly sampled between $-5$ and $25$ $\mathrm{dB}$ .

 Because of its adaptive nature, the online variants of \ac{WPE} have an initialization time $L_i$ (typically $4\mathrm{s}$ with the hyperparameters described below). The utterances are therefore cut into segments of length $L_i$, and the first segment is not used by the end-to-end training procedure, as described in \cite{Lemercier2022, Lemercier2022b}.

Target datapoints are obtained by convolving the first $40$~ms of the \ac{RIR} with the dry speech utterance, simulating the direct path and early reflections, which are beneficial for hearing-aided and normal-hearing listeners \cite{Bradley2003}.

$400$,$ 100$, $60$ \acp{RIR} and $20000$, $5000$, $3000$ clean utterances and noisy excerpts are used for training, validation and testing respectively. Ultimately, each training set consists of approximately $55$ hours of speech data sampled at $16$~kHz.

\newcommand\w{.35\textwidth}
\newcommand\h{.35\textwidth}
\newcommand\xw{0.4}
\newcommand\bw{4.5pt}
\newcommand\xs{.035\textwidth}
\newcommand\ys{-14em}
\newcommand\lxs{0.13\textwidth}

\pgfplotstableread[col sep = semicolon]{data/eval_summary_by_snr.csv}\datatable

\vspace{-0.3cm}
\begin{figure*}[h]
\centering

\begin{tikzpicture}[
    scale=0.855,
    transform shape,
]
\centering

\begin{axis}[
title={$\Delta$PESQ},
name=PESQ,
width=\w,
height=\h,
ybar=0,
ytick={0.0,0.2,...,1.0},
ymin=0.0, ymax=1.1,
bar width=\bw,
major tick length=0pt,
ytick style={
    /pgfplots/major tick length=0.5mm,
},
ymajorgrids=true,
symbolic x coords={snrall, snr1, snr2, snr3},
xticklabels={$\mathrm{SNR} \in [-5;5]$, $\mathrm{SNR} \in [5;15]$, $\mathrm{SNR} \in [15;25]$},
xticklabel style = {font=\small, rotate=45},
xtick=data,
typeset ticklabels with strut,
enlarge x limits=\xw,
mark=none,
error bars/y dir=plus,
error bars/y explicit
]

\addplot+[orange!30!black,fill=orange!10!yellow,mark=none]
table [x={iSNR}, y={PESQimp}, y error={PESQconf95}, col sep=semicolon,
restrict expr to domain={\coordindex}{25:27}] {\datatable}; 

\addplot+[blue!30!black,fill=blue!60!white,mark=none]
table [x={iSNR}, y={PESQimp}, y error={PESQconf95}, col sep=semicolon,
restrict expr to domain={\coordindex}{5:7}] {\datatable}; 

\addplot+[red!30!black,fill=red!45!white,mark=none]
table [x={iSNR}, y={PESQimp}, y error={PESQconf95}, col sep=semicolon,
restrict expr to domain={\coordindex}{17:19}] {\datatable}; 

\addplot+[orange!30!black,fill=orange!10!yellow,mark=none,postaction={pattern=north east lines}]
table [x={iSNR}, y={PESQimp}, y error={PESQconf95}, col sep=semicolon,
restrict expr to domain={\coordindex}{21:23}] {\datatable}; 

\addplot+[blue!30!black,fill=blue!60!white,mark=none,postaction={pattern=north east lines}]
table [x={iSNR}, y={PESQimp}, y error={PESQconf95}, col sep=semicolon,
restrict expr to domain={\coordindex}{1:3}] {\datatable}; 

\addplot+[red!30!black,fill=red!45!white,mark=none,postaction={pattern=north east lines}]
table [x={iSNR}, y={PESQimp}, y error={PESQconf95}, col sep=semicolon,
restrict expr to domain={\coordindex}{13:15}] {\datatable}; 

\end{axis}

\begin{axis}[
title={$\Delta$SDR},
name=sdr,
at={(PESQ.south east)},
width=\w,
height=\h,
xshift=\xs,
ybar=0,
ytick={0, 1, ..., 7},
ymin=0, ymax=7.5,
bar width=\bw,
major tick length=0pt,
ytick style={
    /pgfplots/major tick length=0.5mm,
},%
ymajorgrids=true,
symbolic x coords={snrall, snr1, snr2, snr3},
xticklabels={$\mathrm{SNR} \in [-5;5]$, $\mathrm{SNR} \in [5;15]$, $\mathrm{SNR} \in [15;25]$},
xticklabel style = {font=\small, rotate=45},
xtick=data,
typeset ticklabels with strut,
enlarge x limits=\xw,
mark=none,
error bars/y dir=plus,
error bars/y explicit,
legend style={
    at={(xticklabel cs:.5)},            %
    anchor=north,
    xshift=\lxs,
    /tikz/every even column/.append style={
        column sep=.5cm,
    },
    {nodes={scale=1.0, transform shape}}
},
legend columns=6,
]

\addplot+[orange!30!black,fill=orange!10!yellow,mark=none]
table [x={iSNR}, y={SDRimp}, y error={SDRconf95}, col sep=semicolon,
restrict expr to domain={\coordindex}{25:27}] {\datatable}; 

\addplot+[blue!30!black,fill=blue!60!white,mark=none]
table [x={iSNR}, y={SDRimp}, y error={SDRconf95}, col sep=semicolon,
restrict expr to domain={\coordindex}{5:7}] {\datatable}; 

\addplot+[red!30!black,fill=red!45!white,mark=none]
table [x={iSNR}, y={SDRimp}, y error={SDRconf95}, col sep=semicolon,
restrict expr to domain={\coordindex}{17:19}] {\datatable}; 

\addplot+[orange!30!black,fill=orange!10!yellow,mark=none,postaction={pattern=north east lines}]
table [x={iSNR}, y={SDRimp}, y error={SDRconf95}, col sep=semicolon,
restrict expr to domain={\coordindex}{21:23}] {\datatable}; 

\addplot+[blue!30!black,fill=blue!60!white,mark=none,postaction={pattern=north east lines}]
table [x={iSNR}, y={SDRimp}, y error={SDRconf95}, col sep=semicolon,
restrict expr to domain={\coordindex}{1:3}] {\datatable}; 

\addplot+[red!30!black,fill=red!45!white,mark=none,postaction={pattern=north east lines}]
table [x={iSNR}, y={SDRimp}, y error={SDRconf95}, col sep=semicolon,
restrict expr to domain={\coordindex}{13:15}] {\datatable}; 

\legend{Oracle-RLS-WPE, DNN-RLS-WPE, E2E-RLS-WPE, Oracle-KF-WPE, DNN-KF-WPE, E2E-KF-WPE}

\end{axis}

\begin{axis}[
title={$\Delta$SNR},
name=snr,
at={(sdr.south east)},
width=\w,
height=\h,
xshift=\xs,
ybar=0,
ytick={0, 1, ..., 13},
ymin=0, ymax=13.5,
bar width=\bw,
major tick length=0pt,
ytick style={
    /pgfplots/major tick length=0.5mm,
},
ymajorgrids=true,
symbolic x coords={snrall, snr1, snr2, snr3},
xticklabels={$\mathrm{SNR} \in [-5;5]$, $\mathrm{SNR} \in [5;15]$, $\mathrm{SNR} \in [15;25]$},
xticklabel style = {font=\small, rotate=45},
xtick=data,
typeset ticklabels with strut,
enlarge x limits=\xw,
mark=none,
error bars/y dir=plus,
error bars/y explicit
]

\addplot+[orange!30!black,fill=orange!10!yellow,mark=none]
table [x={iSNR}, y={SNRimp}, y error={SNRconf95}, col sep=semicolon,
restrict expr to domain={\coordindex}{25:27}] {\datatable}; 

\addplot+[blue!30!black,fill=blue!60!white,mark=none]
table [x={iSNR}, y={SNRimp}, y error={SNRconf95}, col sep=semicolon,
restrict expr to domain={\coordindex}{5:7}] {\datatable}; 

\addplot+[red!30!black,fill=red!45!white,mark=none]
table [x={iSNR}, y={SNRimp}, y error={SNRconf95}, col sep=semicolon,
restrict expr to domain={\coordindex}{17:19}] {\datatable}; 

\addplot+[orange!30!black,fill=orange!10!yellow,mark=none,postaction={pattern=north east lines}]
table [x={iSNR}, y={SNRimp}, y error={SNRconf95}, col sep=semicolon,
restrict expr to domain={\coordindex}{21:23}] {\datatable}; 

\addplot+[blue!30!black,fill=blue!60!white,mark=none,postaction={pattern=north east lines}]
table [x={iSNR}, y={SNRimp}, y error={SNRconf95}, col sep=semicolon,
restrict expr to domain={\coordindex}{1:3}] {\datatable}; 

\addplot+[red!30!black,fill=red!45!white,mark=none,postaction={pattern=north east lines}]
table [x={iSNR}, y={SNRimp}, y error={SNRconf95}, col sep=semicolon,
restrict expr to domain={\coordindex}{13:15}] {\datatable}; 

\end{axis}

\begin{axis}[
title={$\Delta$ELR},
name=elr,
at={(snr.south east)},
width=\w,
height=\h,
xshift=\xs,
ybar=0,
ytick={0,1,...,9},
ymin=0, ymax=9.5,
bar width=\bw,
major tick length=0pt,
ytick style={
    /pgfplots/major tick length=0.5mm,
},
ymajorgrids=true,
symbolic x coords={snrall, snr1, snr2, snr3},
xticklabels={$\mathrm{SNR} \in [-5; 5]   $, $\mathrm{SNR} \in [5;15]$, $\mathrm{SNR} \in [15;25]$},
xticklabel style = {font=\small, rotate=45},
xtick=data,
typeset ticklabels with strut,
enlarge x limits=\xw,
mark=none,
error bars/y dir=plus,
error bars/y explicit
]

\addplot+[orange!30!black,fill=orange!10!yellow,mark=none]
table [x={iSNR}, y={ELRimp}, y error={ELRconf95}, col sep=semicolon,
restrict expr to domain={\coordindex}{25:27}] {\datatable}; 

\addplot+[blue!30!black,fill=blue!60!white,mark=none]
table [x={iSNR}, y={ELRimp}, y error={ELRconf95}, col sep=semicolon,
restrict expr to domain={\coordindex}{5:7}] {\datatable}; 

\addplot+[red!30!black,fill=red!45!white,mark=none]
table [x={iSNR}, y={ELRimp}, y error={ELRconf95}, col sep=semicolon,
restrict expr to domain={\coordindex}{17:19}] {\datatable}; 

\addplot+[orange!30!black,fill=orange!10!yellow,mark=none,postaction={pattern=north east lines}]
table [x={iSNR}, y={ELRimp}, y error={ELRconf95}, col sep=semicolon,
restrict expr to domain={\coordindex}{21:23}] {\datatable}; 

\addplot+[blue!30!black,fill=blue!60!white,mark=none,postaction={pattern=north east lines}]
table [x={iSNR}, y={ELRimp}, y error={ELRconf95}, col sep=semicolon,
restrict expr to domain={\coordindex}{1:3}] {\datatable}; 

\addplot+[red!30!black,fill=red!45!white,mark=none,postaction={pattern=north east lines}]
table [x={iSNR}, y={ELRimp}, y error={ELRconf95}, col sep=semicolon,
restrict expr to domain={\coordindex}{13:15}] {\datatable};

\end{axis}

\end{tikzpicture}

\caption{\centering Improvements upon noisy reverberant signals. All metrics except PESQ are in $\mathrm{dB}$. Input \acp{SNR} are indicated in $\mathrm{dB}$.}
\label{fig:barplot}
\vspace{-0.5cm}
\end{figure*}
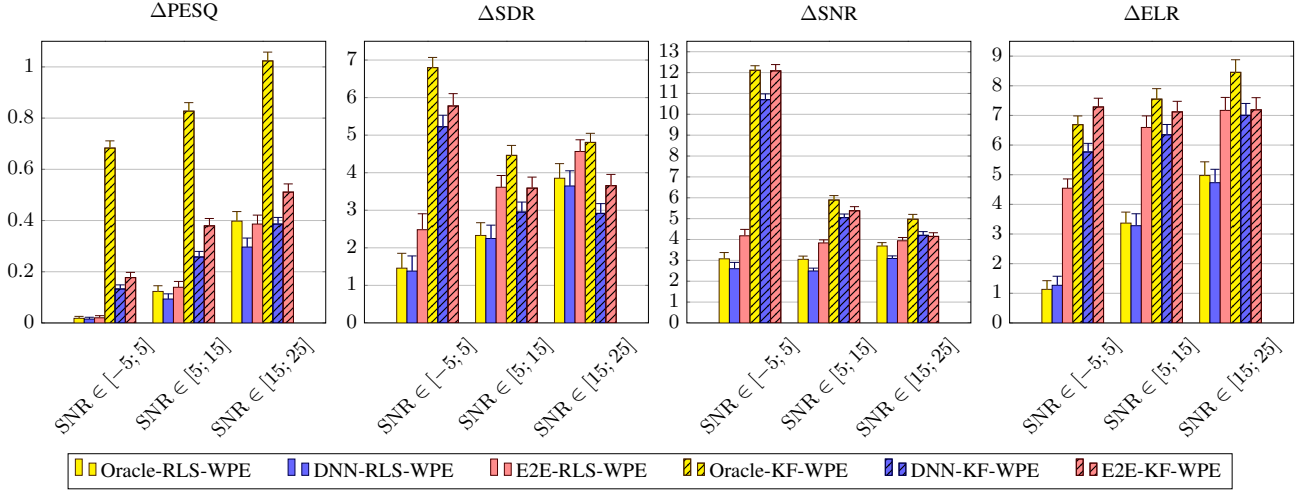

\subsection{Hyperparameter settings}

All approaches are trained with the Adam optimizer using a learning rate of $10^{-4}$ and mini-batches of $128$ segments. The $\mathrm{MaskNet}$ is pre-trained for $300$ epochs, and the $\mathrm{VarNet}$ is trained for $100$ epochs
All networks are optimized with respect to a $L^1$ loss on the spectrogram magnitude.
\ac{DNN} inputs are standardized using the mean and variance of the noisy reverberant distribution approximated by the training set.

The STFT uses a square-rooted Hann window of $32$~ms and a $75$~\% overlap, which yields $F=257$ frequency bins with the corresponding sampling frequency.
The WPE filter length is set to $K=10$ STFT frames (i.e. $80$~ms), the number of channels to $D=2$, the prediction delay to $\Delta=5$ frames (i.e. $40$~ms). The prediction delay value is picked as it experimentally provides optimal evaluation metrics when using the oracle PSD, and matches the target set in the previous subsection. 
When not learnt by $\mathrm{VarNet}$, the filter transition bias is set to $\eta = -35$~$\mathrm{dB}$ as in \cite{Braun2016}, and when it is learnt, the maximal bias is set to $\eta_\mathrm{max} = -30\mathrm{dB}$.

The $\mathrm{MaskNet}$ structure is the same as used in \cite{Lemercier2022, Lemercier2022b}, that is, a single \ac{LSTM} layer with $512$ units followed by a linear output layer with sigmoid activation. 
The $\mathrm{VarNet}$ uses a first linear layer with $F$ output units to fuse the input modalities. A single \ac{LSTM} with $512$ hidden units is then used to model the sequence dynamics, and is followed by a linear layer with $F$ units and sigmoid activation.

We estimate the number of MAC operations per second of our proposed algorithm to $31.4$~GMAC$\cdot\mathrm{s}^{-1}$ at $16$~kHz. This setting allows to obtain a real-time factor---defined as the ratio between the time needed to process an utterance and the length of the utterance---below $0.15$ with all computations performed on Intel(R) Core(TM) i7-9800X CPU.

\subsection{Evaluation metrics}

We evaluate our algorithms on the noisy reverberant test set with the \ac{PESQ} \cite{Rix2001}, output \ac{SDR} and \ac{SNR} \cite{Vincent2006}, as well as the \ac{ELR} \cite{Naylor2011, Yoshioka2011, Lemercier2022}.

\section{Results and discussion}
\label{sec:results}

\subsection{Compared approaches}

We evaluate the RLS-WPE and KF-WPE using the oracle target \ac{PSD} $\lambda$ (\textbf{Oracle-RLS-WPE} and \textbf{Oracle-KF-WPE}).
We then compare the RLS-WPE approach using the DNN-estimated \ac{PSD} (\textbf{DNN-RLS-WPE}, \cite{Heymann2018}) and KF-WPE using the DNN-estimated \ac{PSD} and the filter transition power model \eqref{eq:error} (\textbf{DNN-KF-WPE}, \cite{Braun2016}).
Finally, we evaluate the proposed approach using KF-WPE with DNN-estimated \ac{PSD} and filter transition power \eqref{eq:phi_dnn} (\textbf{E2E-KF-WPE}). 
We also include a RLS-WPE algorithm where the target \ac{PSD} $\lambda$ is estimated by a \ac{DNN} pre-trained using \eqref{eq:pre-training} and fine-tuned end-to-end with \eqref{eq:fine-tuning} (\textbf{E2E-RLS-WPE}, \cite{Lemercier2022}).
Results are displayed in Figure~\ref{fig:barplot}.

\subsection{Oracle experiments}

We first notice that KF-WPE is largely superior to RLS-WPE, if the oracle target \ac{PSD} $\lambda$ is used. In particular, the \ac{SNR} and \ac{ELR} scores of Oracle-KF-WPE indicate that most of the reverberation was removed, although the length of the \ac{WPE} filter only covers $80$~ms, which is largely inferior to the $\T$ times used ($0.4-1.0$~s).
It is also able to remove noise very efficiently, although no denoising mechanism is specified in the signal model.

\subsection{DNN-assisted frameworks}

If the target \ac{PSD} is estimated by $\mathrm{MaskNet}$ without learning the transition power $\phie$, KF-WPE yields agressive dereverberation and denoising performance, thus introducing distortions in the signal. This is confirmed by the high \ac{SNR} and the low \ac{PESQ} and \ac{SDR} scores of DNN-KF-WPE. 

\subsection{End-to-end frameworks}

The proposed E2E-KF-WPE provides superior \ac{PESQ} and \ac{SDR} compared to DNN-KF-WPE, which shows that it is able to circumvent the degrading behaviour of the latter approach. 
Learning the filter transition power $\phie$ helps controlling the adaptation speed as a function of the noise and reverberation condition, thus yielding higher robustness to the estimation errors from $\mathrm{MaskNet}$.
Also, E2E-KF-WPE exhibits high \ac{SNR} and \ac{ELR} scores compared to its RLS-WPE counterparts. This indicates that E2E-KF-WPE is able to significantly improve the dereverberation and denoising performance by using Kalman filtering, especially for low input \acp{SNR}.

We notice in our experiments that the learnt $\phie$ increases with the $\mathrm{T_{60}}$ time and decreases with the input \ac{SNR}, therefore accelerating the adaptation speed as the adversity of the condition intensifies.
We hypothesize that this trend is learnt to compensate for the original model mismatch caused by the WPE filter being too short in comparison to the $\mathrm{T_{60}}$ time on the one hand, and to WPE being noise-agnostic on the other hand.

\section{Conclusion}

We presented a \ac{DNN}-augmented Kalman filtering framework for robust online dereverberation in noisy reverberant environments. Through end-to-end optimization, the DNN estimating the filter transition power is able to control the WPE filter adaptation speed with respect to the noise and reverberation condition.
Our algorithm thus avoids degrading the target signal compared to the original Kalman filtering WPE, while exhibiting stronger dereverberation and denoising power than its RLS-based counterparts.
Future work will be dedicated to inspecting adaptation mechanisms for dynamic acoustics and to the inclusion of suitable denoising approaches in the present framework.

\begin{acronym}
\acro{WPD}{weighted power distortionless}
\acro{DNN}{deep neural network}
\acro{WPE}{weighted prediction error}
\acro{MCLP}{multi-channel linear prediction}
\acro{MVDR}{minimum variance distortionless response}
\acro{GSC}{generalized sidelobe canceller}
\acro{RLS}{recursive least squares}
\acro{EM}{expectation-maximization}
\acro{PSD}{power spectral density}
\acro{ATF}{acoustic transfer function}
\acro{RTF}{relative transfer function}
\acro{RIR}{room impulse response}
\acro{STFT}{short-time Fourier transform}
\acro{SSM}{state-space model}
\acro{SNR}{signal-to-noise ratio}
\acro{LSTM}{long short-term memory}
\acro{HRTF}{Head-related transfer function}
\acro{POLQA}{Perceptual Objectve Listening Quality Analysis}
\acro{PESQ}{Perceptual Evaluation of Listening Quality}
\acro{SDR}{signal-to-distortion ratio}
\acro{ESTOI}{Extended Short-Term Objective Intelligibility}
\acro{ELR}{early-to-late reverberation ratio}
\acro{KF}{Kalman filtering}
\end{acronym}

\bibliographystyle{IEEEtran}

\bibliography{mybib}

\end{document}